\documentstyle[twoside,fleqn,espcrc2]{article}
\def\chiral{{\rm{\bf C}}}
\def\wilson#1{{\rm{\bf B^#1}}}
\def\ham#1{{\rm{\bf H^#1}}}

\newcommand{\AmS}{{\protect\the\textfont2
  A\kern-.1667em\lower.5ex\hbox{M}\kern-.125emS}}

\title{
\vspace{-5.0cm}
\begin{flushright}
{\normalsize DOE/ER/40561-273-INT96-00-134}\\
\vspace{-0.3cm}
{\normalsize UW/PT-96-13}\\
\vspace{-0.3cm}
{\normalsize RU-96-66}\\
\end{flushright}
\vspace*{2.5cm}
Overlap for Majorana--Weyl fermions.}

\author{Rajamani Narayanan\address {Institute for Nuclear Theory, Box 351550
University of Washington, Seattle, WA 98195-1550.}\thanks{Research supported in
part by the DOE under grants \# DE-FG06-91ER40614 and \# DE-FG06-90ER40561.
} and 
Herbert Neuberger\address{Rutgers University, Department of Physics and Astronomy, Piscataway NJ08855.}\thanks{Research supported in part by the
DOE under grant \#DE-FG05-90ER40559.}}

\begin{document}

\begin{abstract}
The power of the overlap formalism is illustrated
by regularizing theories based on Majorana-Weyl
fermions. 
\end{abstract}
\maketitle

The results reported in this contribution were obtained in
collaboration with Patrick Huet \cite{HNN}. 
More than half of the talk will be independent 
of the overlap formalism. 
To present our ideas in the simplest context possible we start directly
from an example. Consider
a $U(1)$ chiral model, defined on a two dimensional torus:
$$
{\cal L} =\sum_{\alpha} \int{\bar\psi_L^\alpha } (\sigma \cdot  (\partial +
iq_L^\alpha A ))\psi_L^\alpha + $$
$$\sum_{\beta} \int{\bar\psi_R^\beta } (\sigma^* \cdot (\partial + 
iq_R^\beta A ))\psi_R^\beta +
{1\over{4g^2}} \int F_{\mu\nu}^{~} F_{\mu\nu}^{~}.$$
The $q_{R,L}^{\alpha , \beta}$  are integers subjected 
to the anomaly cancelation condition
$$
\sum_\alpha ( q_L^\alpha )^2 = \sum_\beta ( q_R^\beta )^2 .$$
Since we are on a torus, 
$$
{1\over{2\pi}} \int F_{12} =0, \pm 1 , \pm 2, \ldots $$ 
All integers can appear, but no other values are allowed.
This implies that the scale of the gauge field is fixed
and consequently the magnitudes of the charges have absolute meaning.
Focus on a single right-handed Weyl fermion of unit charge.
The Dirac-Weyl differential operator is:
$$W_R \equiv \sigma_\mu^* (\partial_\mu + iA_\mu ).$$
Suppose the gauge background carries some topological charge,
$${1\over{2\pi}}\int F_{12} \ne 0 .$$ 
This information alone implies
$$
dim[Ker (W_R )] \ne dim [Ker (W_R^\dagger)].$$ 

If one thinks in matrix terms the differential operator $W_R$ must be
viewed as a rectangular (as opposed to square) matrix. While the matrix is
infinite the distinction still makes sense since it refers to the difference between the number of rows and columns. When $W_R$ is replaced by $W_R^\dagger$
the rows and columns are interchanged.
Suppose we tried to force, by picking
a lattice regularization (or any other method of truncating the space
$W_R$ acts on)  a square shape on $W_R$. The most likely outcome
of this brute force approximation of $W_R$ would be that the square shape
would be used to accommodate representatives of not one continuum
$W_R$ but two continuum differential operators, one of type $W_R$ and the other
of type $W_R^\dagger$. The two rectangular shapes would combine to fit snuggly into the allotted square. 

In the action $W_R$ appears sandwiched between a $\bar\psi$ and a $\psi$.
A square shape for a regularized $W_R$ implies simply an equal number
of $\bar\psi$'s and $\psi$'s. The previous paragraph indicates that in a
topologically nontrivial background one will most likely
find two sets of Weyl fields (or a larger even number) making up
together a Dirac fermion (or several). 
So in nontrivial topology we are led to expect
(in)famous doubling. 

On the other hand, on a lattice the link variables can be smoothly deformed
to connect ``zero topology'' to ``non--zero topology'' gauge configurations.
Thus doubling is there, no matter what the background is.

Of course, doubling is an old story; what we wanted to show is that 
as long as one keeps an equal number of $\bar\psi$ and $\psi$ Grassmann
integration variables on the lattice the number of physical
fermions is likely to be doubled. This cannot be fixed by simply including
one $\bar\psi$ more than the $\psi$'s, say; 
the number of $\bar\psi$ excess we need is background
dependent. It looks like we are in trouble because we cannot 
write down a functional integral as long as we don't know how many integration
variables we have to use. Clearly this story hinges on us being able
to tell apart a $\bar\psi$ from a $\psi$. This is related to being able
to distinguish a particle from its anti-particle. But, we could
use Majorana-Weyl fermions instead, and in this basis the differences
between particles and anti-particles are not notationally apparent.

Let us change variables from one Weyl fermion, $\psi_L$ (left-handed this time)
to two Majorana-Weyl fermions:
$$
\bar\psi_L = {{\xi + i\eta}\over \sqrt{2}}~~~\psi_L ={{\xi -i\eta}\over
\sqrt{2}}$$
$$    
\int \bar\psi_L \sigma\cdot(\partial + i A)\psi_L = $$
$$
{1\over 2}\int \pmatrix{\xi,\eta\cr} 
\underbrace{ \sigma\cdot \left [ \partial +\pmatrix{0&A\cr
-A & 0 \cr } \right ]}_{W_{mL}} \pmatrix {\xi\cr\eta\cr}.$$
We get an $SO(2)$ doublet interacting with an $SO(2)$
gauge field. However, now, $W_{mL}^\dagger =- W_{mL}^*$ and 
$dim[Ker (W_{mL} )]= dim [Ker (W_{mL}^\dagger)] {\rm !}$    
It would seem that $W_{mL}$ can be represented by a square and our
problem went away by changing notation, but this
cannot be true. The fermion number conservation in the
classical Lagrangian has just been obscured by the new notation.

The above exercise indicates that the argument we used to
convince ourselves that an indefinite number of $\bar\psi$,$\psi$
integration variables is needed might not be that general. In
particular, what would happen 
if we had some other irreducible Majorana-Weyl multiplet in a (real)
representation of some gauge group ? We shall see that even then 
there exists an argument forcing us to look for a
representation of the functional integral containing an indefinite
number of fermion fields.

The basic point is that the parity of the number
of zero modes of $W_{mL}$ is a topological
invariant. Somewhere in the vast index literature
this mod(2) index should be mentioned, but we have been unable to locate a 
specific source. We can therefore refer only to 
our own paper \cite{HNN}.  
$Ind \equiv (-1)^{dim[Ker (W_{mL} )]}$ is a topological invariant
essentially  because $W_{mL}$ is skew symmetric. If we again visualize 
the differential operator  $W_{mL}$ as a matrix the latter would be
antisymmetric. The mod(2) index simply tells us whether the matrix
has an odd or even dimension. This makes sense even for
an infinite matrix because the rank of an antisymmetric matrix is always
even, so the parity of its dimension is the parity of the dimension of
the kernel of the operator and the kernel is finite dimensional. 
The index monitors whether the number of Grassmann 
integration variables is even or odd. This parity depends on the gauge field.
So we again face a difficulty because we can't
decide a priori how many Grassmann variables there are.

In the continuum, for a multiplet of Weyl fermions carrying a real
representation, the fermion integral factorizes for any background 
into two isomorphic Majorana multiplets: The 
Majorana-Weyl determinant becomes the fourth root 
of the corresponding Dirac
determinant. 
$$       
\int \bar\psi_L \sigma\cdot(\partial + i {\underline A}
)\psi_L = $$
$$
{1\over 2}\int \pmatrix{\xi,\eta\cr} 
\underbrace{\left [ \sigma\cdot (\partial +i{\underline A})
\pmatrix{1& 0\cr
0 & 1 \cr } \right ]}_{W_{mL}= MW \otimes {\bf 1}_2} \pmatrix {\xi\cr\eta\cr}.$$
The above works because ${\underline A}=-{\underline A}^t$.

The overlap exactly preserves this 
factorization at the regularized level. This is
not entirely trivial since the overlap uses operators of Dirac type to
represent Weyl fermions.
The factorization in the overlap is proven by carrying out a
canonical Bogolyubov transformation. 

We first review the Weyl overlap \cite{NN}:
The exponent of the effective action is given by
$$
z^{\rm lat}_w [U_\mu ] = {}_U\!\! <L-|L+>_{U},$$
where the two states are ground states of two Hamiltonians, each
describing a system of non-interacting fermions:
$$
{\cal H}^\pm =\sum_{x\alpha,y\beta} a^\dagger_{x,\alpha }
\ham\pm(x\alpha,y\beta; U)
a_{y,\beta }.$$
The single particle Hamiltonians are parametrically dependent on the 
gauge fields:
$$
\ham\pm =\pmatrix {\wilson\pm &\chiral\cr 
\chiral^\dagger&-\wilson\pm},\nonumber$$
$$
\chiral(x,y) ={1\over 2}
\sum_\mu \sigma_\mu (\delta_{y,x+\mu}U_\mu (x) ~- ~
\delta_{x,y+\mu}U_\mu^t (y)),\nonumber$$
$$
\wilson\pm (x,y) = ~\pm~ m\delta_{x,y}+\nonumber$$
$$
{1\over 2}\sum_\mu
(2\delta_{x,y}~ - ~\delta_{y,x+\mu}U_\mu (x)~ - ~
\delta_{x,y+\mu}U_\mu^t (y)).
$$  

To go to the Majorana representation we define the transformation:
$$a_1 = {{\alpha_1 - i\beta_1 }\over\sqrt{2}},~~ a_2 =
{{\alpha_2 - i\beta_2 }\over\sqrt{2}}\nonumber$$
$$\alpha_1=\alpha_2^\dagger~~~\beta_1=\beta_2^\dagger .$$
In terms of the new creation/annihilation
operators we obtain:
$${\cal H}^\pm = {1\over 2} \alpha^\dagger \ham\pm \alpha +
{1\over 2} \beta^\dagger \ham\pm \beta .$$ 
Both Hamiltonians split into two terms which commute with
each other and act in separate subspaces. The split into
subspaces is kinematical.
Therefore the overlap factorizes and so would any 
fermionic expectation values. Hence, we can restrict our
attention to only one of the two terms. This reduction
leads us to the overlap formulation of a system containing a single
Majorana-Weyl multiplet. The overlap will be between the ground states
of the Hamiltonians 
$$
{\cal H}^\pm_{mw} = {1\over 2} \Gamma~ \ham\pm_{mw}~ \Gamma ,$$
where the $\Gamma$ operators are hermitian versions of fermionic
creation/annihilation operators: 
$$\Gamma_1 = {{\alpha_1+\alpha_1^\dagger}\over\sqrt{2}}~~~~
\Gamma_2 = {{\alpha_1-\alpha_1^\dagger}\over{i\sqrt{2}}}.$$
The mixing of creation with annihilation operators reflects
the indistinguishableness of Majorana particles from their
anti-particles. The single particle Hamiltonians are real. 

Analogously to the Weyl case, the overlap preserves 
the topological properties of the differential operator it represents.
The regularized version of the mod(2) index is obtained as follows:
The Hamiltonians are operators in 
a Clifford Algebra on a space ${\bf V}$.
In this Algebra,
$
\Gamma_\infty \equiv \prod (\Gamma ),~~\Gamma_\infty^2\propto {\bf 1}
,~~[\Gamma_\infty , {\cal H}]=0$. One can split the space into two subspaces
$
{\bf V}={\bf V}_+  \oplus {\bf V}_-$ that 
are invariant under  the
``chirality'' operator: 
$\Gamma_\infty ({\bf V}_\pm ) = {\bf V}_\pm$. 
All the eigenstates of the Hamiltonians we are dealing with
can be chosen to be entirely
contained in either subspace. 

Now the
source of the index $Ind$ becomes evident: 
$\Gamma_\infty  |L\pm >_U = \zeta |L\pm>_U \Rightarrow Ind=1$,
$\Gamma_\infty  |L\pm >_U =\pm \zeta |L\pm>_U \Rightarrow Ind=-1$.
$\zeta$ is some fixed constant of unit modulus. 
It is also clear that when the index
is $-1$ the chiral Majorana determinant is exactly zero. This
exact ``zero mode'' can be ``soaked up'' by the insertion of a
string containing an odd number of $\Gamma$'s. 

More explicitly, 
the index is given by the determinant of an orthogonal matrix $O$.
This matrix is such that
$O\ham+ O^T$ can be smoothly deformed to  $\ham-$ 
by varying the mass parameter
$m$. Only an even number of generic 
eigenvalue-crossings should occur throughout the deformation. 
One way to find an orthogonal matrix with the above property
is to rotate each $\ham\pm$ to a common canonical form with the
help of two corresponding 
orthogonal matrices $O^\pm$. Then, define $O$ by $O=O^+ O^-$.

\vfill\eject

\end{document}